\documentclass[12pt]{article}

\usepackage[utf8]{inputenc}
\usepackage{amsmath}
\usepackage{graphicx}
\usepackage{xspace}
\usepackage{url}

\usepackage{natbib}

\newcommand{\secref}[1]{\mbox{Section$\,$\ref{#1}}}

\newcommand{\figref}[1]{\mbox{Figure~\ref{#1}}}
\newcommand{\tabref}[1]{\mbox{Table~\ref{#1}}}

\newcommand{\ie}{{\rm i.e.}\xspace}

\newcommand{\etc}{{\rm etc.}\xspace}
\newcommand{\cd}{\,|\,}

\title{A clustering algorithm for the single cell analysis of mixtures}

\author{{Robert G.  Cowell}} 
\date{\today}
\begin{document}
\maketitle
	
\begin{abstract}
A probabilistic clustering algorithm is proposed for the  analysis of forensic DNA mixtures in which individual cells are isolated and short tandem repeats are amplified using the polymerase chain reaction 
to generate single cell electropherograms. The task of the algorithm is to use the peak height information in the electropherograms to  group the cells according to their contributors. Using a recently developed experimental set of individual cell electropherograms, a large set of simulations shows that the proposed clustering algorithm  has excellent performance in correctly grouping  single cells, and for assigning likelihood ratios for persons of interest (of known genotype). 

\end{abstract}


\section{Introduction}
\label{sec:introduction}

In the last decade or so,  probabilistic genotyping for the  computational analysis of forensic short tandem repeat (STR) DNA mixtures  has become  a routine tool, with its use broadly accepted in courts as valid for interpreting DNA mixtures. Over this time such models have become more refined, by including features such as (i) population substructure corrections; (ii) stutter peaks, including double backward and single forward stutters; (iii) ability to treat multiple replicates from a DNA sample (iv) ability to combine peak height evidence from  more than one sample; (v) ability to combined peak height evidence from amplifications  using  more than one kit; (vi) ability to include possible familial relatedness among individuals.  However, with the inclusion of each of these  factors, the computation burden increases significantly with the number of contributors (NoC), an integer that typically needs to be set in order to carry out computations. As the number of contributors increases, it becomes harder to analyze mixtures, not just because of the computation burden, but also because the number of allele peaks seen will tend to increase, and also there is an increasing risk of overlap of genetic profiles that can be difficult to detect and prise apart. In addition, it becomes harder to distinguish the untyped persons by the estimates of their DNA contribution to the mixture, which is the main driver of analysis in probabilistic genotyping.

To overcome these difficulties, a novel procedure has been proposed which uses new technology to isolate individual DNA cells from a DNA mixture, and to amplify them separately.
 Thus, from a  DNA mixture sample, a large number of single cell electropherograms (scEPGs) may be made in the laboratory. Crucially, unlike traditional mixture analysis, each scEPG has only one contributor, and the amount of DNA is known (being just  one cell). 
  For an up-to-date review of the field see
 \cite{huffman2023single}; see also \cite{huffman2023str} for an application to Y-STR loci mixtures.
 
 In principal then, in an ideal world, the genetic profile of the cell for each scEPG can be identified, and where two or more  scEPGS lead to the same cell profile, we can assume that they come from the same person (ignoring identical twins, triplets \etc, all contributing to the mixture). Hence, from a set of  scEPGs, a partition into disjoint subsets of scEPGs may be made, in which all scEPGs in each subset have the same cell profile as their contributor, and scEPGs in any two groups have distinct profiles for the cells' DNA.

Unfortunately we do not live in such an ideal world.  Random variations occur in the amplification process, with wider variation than for bulk mixtures.  This is particularly pronounced in  stutter peaks, which can exceed the parent peaks in height. In addition, but more importantly, dropout of alleles occurs with  high frequency compared with usual mixture analysis, so that a heterozygote can appear as a homozygote on a locus, and even both alleles of a locus can drop out. This is especially acute if the mixture suffers from DNA degradation.  Hence some form of probabilistic analysis becomes important. This paper presents one possible way ahead for such analyses.

There are two main interests in the outcomes of such analyses, just as for bulk mixture analysis. Suppose that we have, connected with the investigation  that a mixture arises from, a set of persons whose profiles are known; denote this set by $\cal K$. Then we may want to know of the mixture:

\begin{enumerate}
	\item Are one or more specified persons from the set $\cal K$ in the mixture (the Persons Of Interest, or POIs)?
	\item How many contributors to the mixture are there, and what are the genetic profiles of the contributors?
\end{enumerate}

For the first of these, the court may require a likelihood ratio of the hypothesis that a  POI is a contributor to a mixture, versus (typically) a random person from the population in place of the POI is a contributor to the mixture. 

For the second,  a listing of the possible profiles of the mixture contributors, typically ranked by their likelihood given the cell EPGs (and the profiles of any untyped individuals) is usually required, for use in a DNA database search of previous offenders.

In an exploration of the application of this approach, 
\cite{duffy2023evidentiary} describe the laboratory preparation of a large number of 
scEPG DNA profiles using the Globalfiler\texttrademark\  STR assay \citep{scientific2015globalfiler}, generated from DNA samples (not mixtures) of several persons of known DNA profile. 
\cite{duffy2023evidentiary} also describe the random generation of synthetic mixtures using these scEPGs, and an analysis of the synthetic mixtures.
The scEPGs were split into two groups, a training set of 1420 scEPGs from 43 individuals that were used to calibrate their probability models for likelihoods evaluation, and a test set of 643 scEPGs of 
good quality (each having at least 15 large allelic peaks).
None of the DNA of  individuals in the training set was used  in the testing set.

The synthetic mixtures, called admixtures,  were  formed by taking  random subsets of the testing set scEPGs for which, unlike real casework  mixtures, the profile of the person was known for each scEPG. Thus the contributor profiles and their mixture ratios to each synthetic mixture was known. The admixtures generated had between 2 and 5 contributors, with total cell numbers on each mixture ranging from between 17 to 75 cells, in a variety of mixture proportions. The  numbers are summarized in 
\tabref{tab:design} (figures taken from \cite{mulcahy2023forensic}).

\begin{table}[htbp]
\caption{Breakdown of the number of contributors (NoC), mixture ratios and cell numbers of the simulated mixtures.}
\begin{center}
\begin{tabular}{cll}\hline
NoC & Mixture Ratio & Number of scEPGs\\ \hline
2 & 1:1 & 15,15  \\
2 & 1:7.5 & 2,15  \\
3 & 1:1:1 & 15,15,15  \\
3 & 1:5:7.5 & 2,10,15  \\
3 & 1:1:7.5 & 2,2,15  \\
4 & 1:1:1:1 & 15,15,15,15  \\
4 & 1:5:5:7.5& 2,10,10,15  \\
4 & 1:1:1:7.5 & 2,2,2,15  \\
5 & 1:1:1:1:1 & 15,15,15,15,15  \\
5 & 1:5:5:7.5:7.5 & 2,10,10,15,15  \\
5 & 1:1:1:1:7.5 & 2,2,2,2,15  \\\hline
\end{tabular}
\end{center}
\label{tab:design}
\end{table}

Now consider a particular random admixture. If we neither use the knowledge of the number of contributors, nor the number of cells from each individual, then we may consider the following types of analyses that could be carried out.

\begin{enumerate}
\item Infer the number of contributors (NoC).
\item Partition the scEPGs into clusters, so that all the cells in one cluster come from a single person, and infer the genotype of each such person.
\item Using a (peak height) probability model, evaluate the likelihood of seeing the peak height profiles in the scEPGs.

\item
 Additionally, if there is a person if interest (POI) of known profile, find the likelihood ratio that (a person matching ) the POI is a contributor to one or more cells of the mixture versus, the alternative that they are not.
\end{enumerate}

\cite{duffy2023evidentiary} tackle the partitioning (2) of the  scEPGs into clusters  by using \textit{Model Based Clustering} (MBC), a machine learning pattern matching algorithm. The number of clusters in the partition is identified with the NoC. One can then use the knowledge of the true contributor to each cell to determine (a) if the NoC is correct, and (b) if the cells in each cluster  indeed comes from a single person. In this way the efficacy of the clustering algorithm may be judged.

Having found the partition, probabilistic genotyping software may be used to find for each cluster the likelihoods of the observed peak heights in the scEPGs of each cluster -- assuming that the each cluster is generated from one person's profile. Then,  multiplying the likelihoods across the clusters yields the likelihood for the partition. Additionally, for each cluster probabilistic genotyping software can be used to infer the most likely profile of the contributor, and this can be compared to the true (known) contributor, (assuming that the scEPGs of a cluster have been correctly identified as coming from one person).

\cite{duffy2023evidentiary} also present formulae for the likelihood ratio of point (4) above, but the correctness of their formula depends upon the MBC algorithm picking out a correct partitioning of the scEPGs in the mixture. Although their use of the MBC algorithm appears to perform very well, it does not always yield a correct partition.
 
It is arguably  unsatisfactory to use the MBC algorithm to do the clustering, having at its heart one probability model of peaks heights of the scEPGs used in the  pattern matching, and then a  distinct  probability model to evaluate the  scEPG likelihoods given the partition found. A more elegant approach would be to use the same probabilistic peak height model to both 
do the clustering and use it to find likelihoods from the clusters. 

Such an approach was made by \cite{lun2024calculation}, who proposed an algorithm utilising \textit{hierarchical  clustering}, which they called \textit{forensic-aware clustering (FAC)}. This has been applied by 
\cite{qhawe2025}, who compared MBC and FAC  on a set of 630 simulated admixtures. The results were much improved by the use of the latter algorithm. However, one aspect of the hierarchical  clustering algorithm  is that it generates only one partition which must be assumed to be the correct one to apply 
the formula of  \cite{duffy2023evidentiary} to evaluate likelihood ratios in a case having POIs.

In this paper we  present an alternative clustering algorithm that builds up candidate partitions cell-by-cell.  It is applicable for the cases either  with and without POIs. The algorithms are applied to simulated mixtures, similar to those in \cite{duffy2023evidentiary}, to evaluate their efficacy. The algorithm generates not just one partition, but a number of partitions each of which has a likelihood attached; these can be used to give posterior probabilities of correctness to each generated partition.

The plan of the paper is as follows. The next section discusses some considerations for evaluating the likelihoods that will be used in the simulations. We then present the clustering algorithm, first for the case of no POIs, and then with POIs. Then a large number of admixtures are simulated, some using the design  in \tabref{tab:design}, though not identical to the mixtures  used in \cite{duffy2023evidentiary}. The results from those following the  design  in \tabref{tab:design} are compared to  those in \cite{duffy2023evidentiary} who used MBC. A comparison to hierarchical  clustering algorithm  on mixtures without POIs is given, both from a theoretical  perspective and a comparison of simulated mixtures. Finally results of  simulations involving POIs are given, followed by  brief conclusions.

\section{Likelihood considerations}

\subsection{Likelihood evaluation}
Suppose that we have a set of scEPGs of cells which are assumed to  have come from a single DNA sample, possibly a mixture of several persons' DNA, and that all the cells are amplified using the same PCR kit and protocols. We shall assume that there are no known DNA profiles under consideration, i.e., $\cal K = \emptyset$, and for simplicity we shall assume that there is no population substructure correction to make ($\theta = 0$).

From a computational side, we shall assume that for any locus $m$ and genotype $g^m$ we may calculate the genotype probability $p(g^m)$.  We also have a model to evaluate the likelihood for a peak-height profile $e^m$ on the locus $m$ given the genotype $g^m$.
Let $e = \cup_m e^m$ denote the peak height profiles on all the loci, and similarly  $g = \cup_m g^m$ the genotype over the loci.

The genotype probability and conditional probability (or probability density) of the scEPG $e$ given it arises from the amplification of the single cell with profile $g$, are  given by, (using the independence of loci),

\begin{align*}
	p(g) &=\prod_m p(g^m)\\
	p(e\vert g)&=\prod_m p(e^m\vert g^m)\\
\end{align*}
where we also assume that $e^m$  depends only on the genetic profile
$g^m$ and does not depend on any other locus profile $g^j: j \ne m$.

Let us assume that we have $n$ cell peak height profiles $E_1,\ldots, E_n$ (so that $E_i = \cup_m\{ E_i^m\}$), and denote the totality of the profiles by $E$.

Now the ground truth is that there is some number $K$ of contributors having distinct genotypes $g_1,g_2,\ldots,g_K$ such that the $n$ cell profiles
are partitioned into $K$ clusters $C = \{c_1,\dots,c_K\}$ with contributor $k$ giving rise to all of the EPGs in the cluster $c_k$.

Now we don't know the number $K$, we don't know the  clusters $C$, and we don't know the genotypes $G=\{g_1,g_2,\ldots,g_K\}$ to be associated with the clusters.
Our task is to infer these, using  a probabilistic approach. 

We think of this as an exercise in model selection, or as model averaging, in which each model is a partition of the set of scEPGs into clusters. We have a likelihood score associated with each cluster, and a way of combining the scores to give one for the cluster partition. We then choose the model, or clustering, that optimizes the score (maximizes the likelihood). Alternatively, we may calculate averages based on all of the partitions found, using the score of each cluster, as in Bayesian model averaging, which will require a prior over the partitions.

In a frequentist approach the score for a cluster $c_i$ will be the likelihood given by

$$L(E,c_i) = \sum_{g_i} p(g_i)\prod_{E_j\in c_i} p(E_j\vert g_i),$$
with the overall score the likelihood

$$ L(E,C) = \prod_{c_i \in C} L(E,c_i).$$

Alternatively we may take the log-likelihood as the overall score:

$$ LL(E,C) = \log L(E,C) = \sum_{c_i \in C} \log L(E,c_i)$$

In a Bayesian approach we require a prior over the partitions, so we have $p(C)$ for every partition $C$. If we have likelihoods for every possible partition, we may use the likelihood above to evaluate a posterior distribution

$$p(C\vert E) \propto  L(E,C)p(C)$$
and find the cluster $C$ that maximizes the posterior score; alternatively we may use the posterior distribution to carry out model averaging. In this paper we use a uniform distribution over partitions, $p(C) = \mbox{constant}$, for which  the posterior distribution will be simply the normalized likelihood. This is distinct from the proposal of \cite{lun2024calculation}, who base their FAC algorithm on a uniform prior over the number of contributors. We shall come back to this point in \secref{sec:aggloclust}.

As an example, consider the NoC. Associated with each partition into clusters  $C$ is the total number of clusters $n_C$. Hence, if we have the posterior distribution 
$p(C\vert E)$ we may evaluate (by marginalization) the posterior distribution $p(NoC\vert E) $ by summing the terms of the distribution for which the clusters $C$ have exactly $k$ non-empty sets:

$$p(NoC\vert E) = \sum_{C: \vert C\vert = NoC}p(C\vert E). $$

When there are POIs included, then the specification of priors is a little more complex, as we shall see later.

\subsection{Accounting for population substructure}

Given a partition, we can evaluate the likelihood $L(c_i)$ of a cluster by evaluating the likelihoods of the cluster for each locus of interest, and then multiplying these together, so this gives some savings on the computational complexity. However, when we multiply the various $L(c_i)$ together over the different clusters, we are assuming no population substructure correction. In other words, to include population substructure correction \textit{precisely} would make evaluations much more complicated and we lose the simple product rule. An approximation would be to apply it to each cluster separately, which is what we shall do. 

To see the numerical difference consider just two clusters, and hence two untyped persons, and a locus $m$. Consider a term  where the genotype is homozygous with profile  $(a,a)$ in each cluster. If $\theta$ denotes the population substructure parameter, then the precise correction would yield the joint probability (with $p=p(a)$ the population frequency for the allele $a$)

$$P((a,a),(a,a))=\frac{p(\theta+p(1-\theta))(2\theta+p(1-\theta))(3\theta + p(1-\theta))}{(1+\theta)(1+2\theta)}$$
which reduces to $p^4$ when $\theta = 0$.

Alternatively an approximation would be to use the correction to second order in each person separately, thus to use for each person the probability

$$ P((a,a)) = p(\theta+p(1-\theta))$$
which is applied to each cluster evaluation separately, and then multiplied together for the joint genotype probability. The difference is given by subtracting the square of this expression from the previous one, yielding

$$\frac{p (p (1-\theta)+\theta) (p (1-\theta)+2 \theta) (p (1-\theta)+3 \theta)}{(\theta+1) (2 \theta+1)}-p^2 (p (1-\theta)+\theta)^2$$

For $\theta=0.02$ the difference plotted out as a function of $p$ gives the following figure, with $p$ the horizontal scale.\vspace{12pt}

\begin{center}
\includegraphics[scale=0.65]{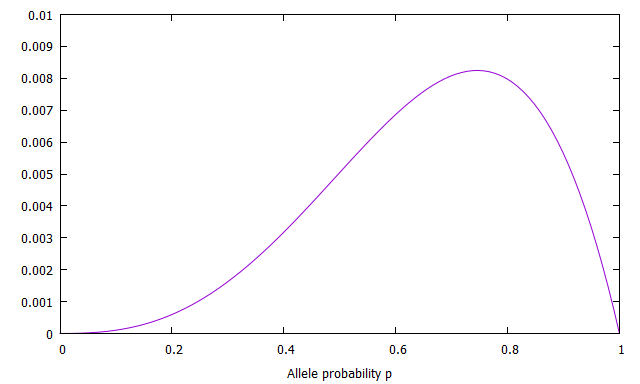}
\end{center}
	
\vspace{12pt}
We see the largest difference of around 0.008 is at around $p=0.8$ for which $p^4 = 0.4096$. For $p=0.2$, for which $p^4=0.0016$, the difference is approximately $0.00059$. 
Alterrnatively, looking at the ratio

$$\frac{p (p (1-\theta)+\theta) (p (1-\theta)+2 \theta) (p (1-\theta)+3 \theta)}{(\theta+1) (2 \theta+1)} \bigg/ p^2 (p (1-\theta)+\theta)^2$$
plotted in the following figure
\begin{center}
	\includegraphics[scale=0.65]{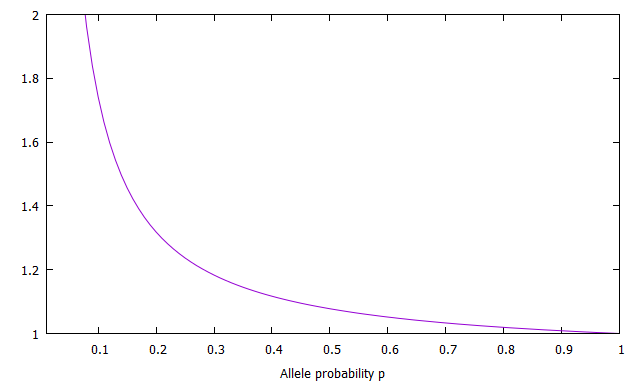}
\end{center}

\noindent
we can see the  ratio rising for small $p$.

 If we consider  the profiles $(a,b)$  and $(a,b)$, and denote $p = p(a)$ and $q = p(b)$, then the joint profile probability
will be

$$ P((a,b),(a,b)) = \frac{4 \left(\left(1-\theta \right) p+\theta \right) \left(1-\theta \right) p \left(\left(1-\theta \right) q+\theta \right) q}{\left(1+2 \theta \right) \left(1+\theta \right)}$$
whereas

$$ P((a,b)) P((a,b)) = 4(1 - \theta)^2p^2q^2$$

For 2-D plotting convenience, set $p=q$; the ratio becomes

$$ \frac{\left(\left(1-\theta \right) p+\theta \right)^{2}}{\left(1-\theta \right) p^{2} \left(1+2 \theta \right) \left(1+\theta \right)}$$ 
and is plotted here where because we have set $p=q$ we must have $p \le 0.5$:

\begin{center}
	\includegraphics[scale=0.65]{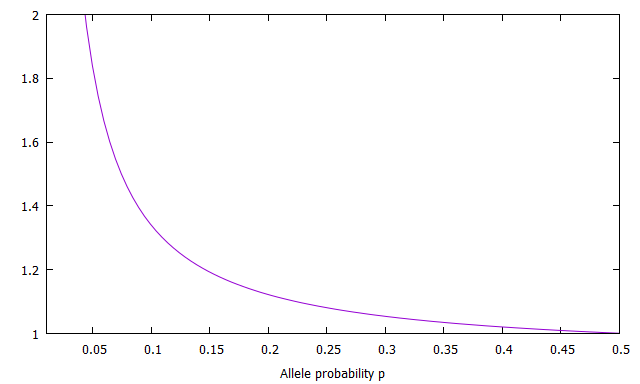}
\end{center}

Again we see an values above 1, increasing rapidly for lower $p$ values.

However, many of the cross terms of profiles for the two individuals will have no alleles in common: $(a,b), (c,d)$.
In this case we have
$$P((a,b),(c,d)) = \frac{4 p_ap_bp_cp_d (1-\theta)^3}{(1+\theta)(1 + 2\theta)},$$
whilst
$$P((a,b))P((c,d)) = 4 p_ap_bp_cp_d (1-\theta)^2,$$
leading to the ratio
$$ \frac{P((a,b),(c,d))}{P((a,b))P((c,d)) } = \frac{1-\theta}{(1+\theta)(1 + 2\theta)},$$
which for $\theta = 0.2$ gives a value of approximately 0.924, irrespective of the allele frequencies, a decrease for all such terms. These will counteract to some extent the increases in the cases considered earlier.

Thus in using the approximation of applying $\theta$ corrections individually to clusters in a partition, there will be terms where the likelihood should be given a larger weight and others a lower weight. Overall it is hard to say what the final result will be, but as there may be expected to be a fair amount of cancellation of corrections, and given that the value of $\theta$ is not precisely known, it would seem that using the above approximation is acceptable, and we use it. (This assumed cancellation is similar to the use of random phase approximation in physics.)

\subsection{The size of partition space}
Given a set of  $n$ distinct objects (the scEPGs), the number of partitions into non-empty sets (clusters) is given by the Bell numbers $B(n)$\footnote{See https://en.wikipedia.org/wiki/Stirling\_numbers\_of\_the\_second\_kind}, which grow super exponentially.
The sequence of the first twelve Bell numbers is given by
$$1,1,2,5,15,52,203,877,4040,21147,115975,678570,4213597.$$

The number of partitions into $k$ non-empty clusters is given by the Stirling numbers of the second kind, which we denote by $S(n,k)$. \tabref{tab:stirling1}  shows values for small $n$ and $k$ values up to 10.
We have the Bell numbers given by

$$B(n) = \sum_{k=0}^n S(n,k)$$

\begin{table}[htbp]
	\caption{\label{tab:stirling1}Stirling numbers $S(n,k)$ for $1 \le n, k \le 10$. 
		Row sums give the Bell numbers, with the final row sum of the table giving  $B(10) = 115975$.}
		\begin{center}
\begin{tabular}{c|cccccccccc}
	$n \backslash k$& 1&2&3&4&5&6&7&8&9&10\\\hline
1& 1 & 0 & 0 & 0 & 0 & 0 & 0 & 0 & 0 & 0 \\
2& 1 & 1 & 0 & 0 & 0 & 0 & 0 & 0 & 0 & 0 \\
3&1 & 3 & 1 & 0 & 0 & 0 & 0 & 0 & 0 & 0 \\
4&1 & 7 & 6 & 1 & 0 & 0 & 0 & 0 & 0 & 0 \\
5&1 & 15 & 25 & 10 & 1 & 0 & 0 & 0 & 0 & 0 \\
6&1 & 31 & 90 & 65 & 15 & 1 & 0 & 0 & 0 & 0 \\
7&1 & 63 & 301 & 350 & 140 & 21 & 1 & 0 & 0 & 0 \\
8&1 & 127 & 966 & 1701 & 1050 & 266 & 28 & 1 & 0 & 0 \\
9&1 & 255 & 3025 & 7770 & 6951 & 2646 & 462 & 36 & 1 & 0 \\
10&1 & 511 & 9330 & 34105 & 42525 & 22827 & 5880 & 750 & 45 & 1 \\	
\end{tabular}
\end{center}
\end{table}

Now for the case of DNA mixtures, we might expect a typical $n$ value to be
96 or more, with 96 corresponding to the number of simultaneous amplifications possible in a full PCR plate. Hence $B(96)$ would be astronomical (it is approximately $6.68\times 10^{109}$). However, we might expect the number of contributors to be small, so that the $S(n,k)$ for moderate values of $k$ would be more appropriate. However, even for $k=6$ which is around the computational  limit for current probabilistic genotyping mixture software, we have 
$S(15,6)=420,693,273$ already an enormous number, and 
$S(96,6)\approx 7.00 \times 10^{71}$. Even for $k=2$ we have $S(n,2) = 2^{n-1}$ which is approximately $3.96\times  10^{28}$ for $n=96$.

It is therefore probably out of the question that an exhaustive model search or computation of all model scores could be ever carried out, and so some search strategy is required to look for the high likelihood partitions.

\section{Generation of partitions without POIs}

There are two main ways of generating partitions of $n$ distinct objects. One way is to start from a particular partition, say that consisting of a single cluster of the $n$ objects, and then apply a modifying algorithm to generate a new partition. 
The modifying algorithm is repeatedly applied to each newly formed partition to make a new partition, but without re-creating a previously generated partition, until all partitions have been generated. This type of algorithm is not practical for our application.

The other way is to build up the partitions incrementally, and is the method we pursue here.
We now look at doing a search for high scoring cluster partitions. We begin with an exact (exhaustive) evaluation that generates all possible partitions, and then consider ways to get good approximations from this by avoiding what would seem to be the very low scoring partitions.

\subsection{Basic setup and assumptions}

Let us summarize the setup here. We have a set $n$ of scEPGs that have been amplified, taken from a DNA sample which may or may not be a mixture of several persons' DNA. The peaks height information (evidence) in these scEPGs are labelled by $E_i$ with $i\in (1,n)$.
The $n$ cells could have come from one unknown person, $n$ different unknown persons, or any number in between. We do not assume or know of any familial relatedness between these unknown (untyped) individuals.
We assume that the individuals come from the same population. 

The profiles are generated from some kit using a set of STR   loci $L$. We assume that the loci  are all autosomal, with the possible exception of Amelogenin, so that the product rule can be applied.  For each locus $m$ we have a set of alleles $A^m$. We have estimates of the population allele frequencies so that we can find for any genotype $g^m$ on locus $m\in L$ with alleles in $A^m$ the genotype probability $p(g^m)$ (with or without population substructure correction as desired). 

We also have a peak height probability model, for which we can evaluate for each locus $m$ the likelihood $p(E^m_i\vert g^m)$ for any genotype, so that using the product rule we can also evaluate
for each cell

$$p(E_i) = \prod_m \sum_{g^m} p(E^m_i\vert g^m)p(g^m), $$
which uses the product rule over loci. For a non-empty subset of cells $\cal I$ this generalizes to

$$p(\{E_i:i \in {\cal I}\}) = 
\prod_m \sum_{g^m} p(\{E^m_i:i \in {\cal I}\}\vert g^m)p(g^m). $$

The interpretation of this is that all the cells yielding the scEPGs $\{E_i:i \in {\cal I}\}$ in the cluster $\cal I$ come from the same untyped person. If ${\cal I}$ is part of a partition of the cells, say into $k$ parts, as we multiply the values from the expression above obtained from all parts of the partition, then we have the likelihood of the partition given the data in the scEPGs. The ideal aim is to evaluate this likelihood for all possible partitions. 

\subsection{Clustering Algorithm without POIs}
\label{sec:algnopoi}

We propose an algorithm where we generate sets of partitions as we iteratively add individual cells: we call the algorithm the \textbf{Incremental Partitioning Algorithm} (IPA). It works as follows.

We arrange the cells into a linear ordering. We then take the first cell and  evaluate

$$p_1(E_1) = \prod_m \sum_{g^m} p(E^m_1\vert g^m)p(g^m) $$

This is associated with the single partition $(1)$ of the set $\{1\}$, and corresponds the cell coming from an untyped person  $u_1$.

Next we take the second cell. For this we evaluate

$$p_1(E_1, E_2) = \prod_m \sum_{g^m} p(E^m_1, E^m_2\vert g^m)p(g^m) $$

This is associated with the single partition $(1,2)$ of the set $\{1,2\}$, and corresponds to a person  $u_1$ being the  donor of both cells.
We also evaluate

$$p_2(E_1, E_2) = p_1(E_1)p_1(E_2),$$
where

$$p_1(E_2) = \prod_m \sum_{g^m} p(E^m_2\vert g^m)p(g^m). $$

This is associated with the single partition $(1),(2)$ of the set $\{1,2\}$, and corresponds to a person  $u_1$ being the cell donor of  cell 1, and a second untyped person $u_2$ being cell donor of  cell 2.

The  way to think of this is that from the first step we have a single partition for a single person $u_1$ for  a single scEPG. From this we make two possible partitions on adding a second cell, in which either (i) the second cell is also assumed to come from the person $u_1$, or (ii) the second cell comes from another person. This gives two possible partitions of the first two cells, $(1,2)$ and $(1),(2)$. 

Now we can add the third cell. There are three possibilities: (i) This cell also originates from the first person, $u_1$; (ii) the cell originates from the second person $u_2$, (iii) the cell originates from yet another person $u_3$. These three possibilities mean that, from the two  partitions $(1),(2)$ and  $(1,2)$, we may generate the five partitions  as follows:
\begin{align*}
	(1),(2) &\to  (1,3),(2)\\
	(1),(2) &\to  (1),(2,3)\\
	(1),(2) &\to  (1),(2),(3)\\
	(1,2)&\to  (1,2,3)\\
	(1,2)&\to  (1,2),(3)
\end{align*}

We may evaluate the likelihoods for each of these. 
In general, suppose that after $k$ cells we have the $B(k)$ (Bell number) partitions. Then to add the next cell we take each partition, and consider adding the new cell to each of the subsets  of the partition, plus appending the cell as singleton to the partition. If we do this for all the $B(k)$  we generate the set of $B(k+1)$ partitions. We can evaluate the probabilities for the scEPGs for each of these partitions.

However, we have seen that the Bell numbers grow very rapidly. 
Therefore, after having generated the new partitions using the extra cell, we then evaluate their likelihoods and discard all partitions that are lower in likelihood by some factor, say a billion, than the top scoring partition. This removes a great many low scoring partitions, and prevent these from generating further low scoring partitions as more cells are added.
 With this pruned search we generate a more manageable set of high scoring partitions, typically just a few dozen.

So to summarize, starting with the empty partition  we add  cells one by one such that
when cell $k$ has been added we have a set of candidate partitions. We add a new cell by making all possible extensions of the candidate partitions, and calculate the likelihoods for each of them. Of these we retain only those having a high enough likelihood.  Schematically the algorithm is:

\begin{itemize}
	\item Put the first cell in its own cluster, making one partition.
	\item for cell $i=2,3,4\ldots$ in sequence
	\begin{itemize}
		\item Extend the current partitions in all possible ways with cell $i$ to make a new set of partitions.
		\item Score each partition's likelihood (by  the peak height probability model).
		\item Discard partitions with relatively poor likelihood scores.
	\end{itemize}
\end{itemize}

\section{Generation of partitions with POIs}
\label{sec:plp}

The procedure above is readily generalized to deal with having profiles of one or more persons of interest (POI) whose genetic profile is known. To do so we need to introduce the notion of a \textit{Partially Labelled Partition} 

\subsection{Partially labelled partitions}

\subsubsection{A single POI} 
Consider initially that case of a single POI having label $s$. Now in our iterative scheme for the first cell we  have 
 two options to consider: (i) the cell comes from $s$; (ii) it comes from an untyped person $u$. So for one cell we have two ``partitions".  We will call these partitions \textit{Partially Labelled Partitions} (PLPs), where the POIs supply the (optional) labelling. (These do not appear to have been considered in the literature before, but surely have in some guise.)

So with one POI $s$ and a one cell labelled $1$ we have two possible PLPs. One is labelled with $s$ which we denote by $\{(1\cd s)\}$, the other not labelled with $s$  and denoted by $ \{(1)\}$.

Let us add cell number 2. We may extend the PLP  $\{(1\cd s)\}$ in two ways: either cell 2 also comes from $s$, or is from an untyped person. Hence the  two possible PLPs  generated are:

\begin{itemize}
	\item $\{(1,2\cd s)\}$
	\item $\{(1\cd s), (2)\}$
\end{itemize}

We may extend the PLP  $ \{(1)\}$ in three ways: (i) cell 2 comes from the same untyped person as cell 1; (ii) cell 2 comes from another untyped person, or (iii) cell 2 comes from the POI $s$. We may denote these three PLPs as follows:

\begin{itemize}
	\item $\{(1,2)\}$
	\item $\{(1), (2)\}$
	\item $\{(1), (2\cd s)\}$
\end{itemize}

This makes as a total of 5 PLPs with two cells and with one typed person $s$
We see that the PLPs are growing faster in number than the Bell numbers. 
In fact, with the Stirling numbers $S(n,k)$, representing the number of partitions of $n$ objects into $k$ non empty  subsets,  there will be $k$ ways of allocating $s$ to a cluster, and 1 in which we do not label any cluster, so that there are $(1+k)S(n,k)$ PLPs with $k$ clusters.
Hence the total number of PLPs with $n$ cells and 1 POI will be

$$ \sum_{k=1}^n (1+k)S(n,k)$$

\subsubsection{Two POIs}

Given two typed POIs $s_1$ and $s_2$, when we start with the first cell we have three options, in which one of the POIs is associated with the cell, or neither is:

\begin{itemize}
	\item $\{(1\cd s_1)\}$ 
	\item $\{(1\cd s_2)\}$ 
	\item $ \{(1)\}$
\end{itemize}

In adding the second cell, the first of these can be extended in three ways:

\begin{itemize}
	\item $\{(1, 2\cd s_1)\}$ 
	\item $\{(1\cd s_1)\}$  and $\{(2\cd s_2)\}$ 
	\item $\{(1\cd s_1)\}$  and $ \{(2)\}$
\end{itemize}

The second $\{(1\cd s_2)\}$ PLP can similarly be extended to

\begin{itemize}
	\item $\{(1, 2\cd s_2)\}$ 
	\item $\{(1\cd s_2)\}$  and $\{(2\cd s_1)\}$ 
	\item $\{(1\cd s_2)\}$  and $ \{(2)\}$
\end{itemize}

and the third PLP $ \{(1)\}$ as follows

\begin{itemize}
	\item $\{(1)\}$  and $\{(2\cd s_1)\}$ 
	\item $\{(1)\}$  and $\{(2\cd s_2)\}$ 
	\item $\{(1)\}$  and $\{(2)\}$ 
	\item $\{(1, 2\}$ 
\end{itemize}

\noindent
making a total of 10 PLPs in all.

These numbers are again growing faster than the Bell numbers.

\subsubsection{General case of $m$ POIs}

More generally, if we have $m$ POIs and $n$ cells then the total number of PLPs will be

\begin{eqnarray*}
P[n,m] &=& \sum_{j=0}^m\sum_{k=1}^n \binom{m}{j}\binom{k}{j} j!\  S(n,k)\\
&=& B(n) + \sum_{j=1}^m\sum_{k=1}^n \binom{m}{j}\binom{k}{j} j!\  S(n,k) \\
\end{eqnarray*}

This is derived as follows. 
Given $m$ POIs, there are
 there are  $\binom{m}{j}$ ways of choosing  $j$ POIs from them. 
 If we have a partition of $n$ objects into $k$ clusters, then there are 
$\binom{k}{j}$ ways of choosing $j$ of the clusters to label with a POI, and $j!$ permutations of  arranging the $j$ POI labels in these selected clusters. Multiplying these factors with the Stirling number $S(n,k)$ and summing over $j$ and $k$ gives the formula above.
These numbers grow very rapidly, as shown in \tabref{tab:plp}.\footnote{Interestingly the numbers in this table appear in \cite{whitehead1978stirling} dealing with chromatic polynomials.}

\begin{table}[htbp]
	\caption{Numbers of Partially Labelled Partitions on $n$ objects and $m$ labels  $P[n,m]$. The first column with $m=0$ gives the Bell numbers.}
	\begin{center}
		\begin{tabular}{c|rrrrrrrrrr}
		& \multicolumn{8}{c}{$m$}\\  
		$n$ & 0 & 1  & 2&3&4&5&6&7&8\\ \cline{2-10}
		1 & 1 & 2 & 3 & 4 & 5 & 6 & 7 & 8 & 9  \\
		2 & 2 & 5 & 10 & 17 & 26 & 37 & 50 & 65 & 82  \\
		3 & 5 & 15 & 37 & 77 & 141 & 235 & 365 & 537 & 757  \\
		4 & 15 & 52 & 151 & 372 & 799 & 1540 & 2727 & 4516 & 7087  \\
		5 & 52 & 203 & 674 & 1915 & 4736 & 10427 & 20878 & 38699 & 67340  \\
		6 & 203 & 877 & 3263 & 10481 & 29371 & 73013 & 163967 & 338233 & 649931  \\
		\hline
		\end{tabular}
	\end{center}
	\label{tab:plp}
\end{table}

\subsection{Clustering Algorithm with POIs}
\label{sec:algpoi}

We can now discuss the general algorithm of partition searches with POIs.

In general, suppose we have $m$ POIs, then with the first cell we can make $m+1$ PLPs, by assigning one of the POI to the cell (in $m$ ways) or none (corresponding to an untyped person).

Suppose that after $k-1$ cells we have a set of PLPs. Then we can extend each PLP in  ways that depend of the number of clusters that are labelled and  those that are not in the PLP.  Suppose that $j$ of the clusters are labelled and $n$ are not, so the PLP has $n+j$ clusters overall, and we have a new cell $c$ to add. Then it can be added to make new extended PLPs in the following ways:
\begin{itemize}
	\item Make $n+j$ PLPs by adding $c$ to each one of the existing $n+j$ clusters.
	\item Make 1 PLP by adding $c$ as its own cluster (thus introducing a new untyped person)
	\item Make $m-j$ PLPs by adding a new cluster $c$ labelled by the $m-j$ POIs that are NOT already present in the original PLP.
\end{itemize}

This makes a total of  $m+n+1$ new PLPs from the original PLP of $n+j$ clusters. In this way we can generate a  set of new PLPs from the original PLPs. However, to avoid the huge explosion in the number of partitions formed, as for the non POI case, as we build up the PLPs cell by cell we evaluate the PLPs likelihoods and discard those that are very small relative to the highest likelihood PLP. These PLPs and their likelihoods may be used to find likelihood ratios for various pairs of hypotheses. However, there is an additional symmetry factor that must be incorporated in each PLP likelihood before this can be done. The origin and form of this factor is explained in the next section.

\section{Additional symmetry factors for PLP likelihoods}

The main intended use of the algorithm for scoring PLPs in the presence of POIs is to  evaluate likelihood ratios for the presence of the POIs in the mixture(s), for providing evidence in court. The simplest case is that of having a single POI. For this we will need a numerator for the likelihood of the POI and possibly other untyped persons to be in the mixture, and in the denominator the likelihood for untyped persons to be all the contributors to the mixture. The algorithms presented above provide these likelihoods in the absence of POIs, but misses an important factor when POIs are included. 

To understand the origin of this factor, we shall use a Bayesian approach, and look at finding Bayes factors for POIs to be contributors to a mixture. For this we shall need to find posterior probabilities found by averaging the likelihoods using prior probabilities. We shall examine this by adding POIs, starting with the case of having no POIs. We shall assume the use of the algorithms but without any pruning of the partitions as cells are added. We thus end up with all possible  partitions.

\subsection{Marginal likelihood for no POIs}

Suppose we have $n$ scEPGs. There is a total of $B(n) =\sum_{k=1}^n S(n,k)$ possible partitions that can be made with these $n$ scEPGs, each will have a peak height likelihood as described above. Let $\sigma_{n,k:i}$ denote a partition of the $n$ scEPGs into $k$ clusters, its likelihood denoted by $L_{n,k:i}$, with the index $i$ ranging from 1 to $S(n,k)$.
Let $p_{n,k:i}$ denote the prior probability that $\sigma_{n,k:i}$  is the true partition. Then the marginal likelihood is given by

$$L(n)= \sum_{k=1}^n \sum_{i=1}^{S(n,k)} L_{n,k:i}p_{n,k:i},$$
with the constraint

$$\sum_{k=1}^n \sum_{i=1}^{S(n,k)} p_{n,k:i} = 1.$$

In the absence of prior information about which  of the  partitions is the true partition,  a uniform distribution over the partitions is plausible. This would mean that $p_{n,k:i}  = 1/B(n)$ for every allowed $k$ and $i$. Note that this implies the prior probability for the number of contributors to be $j \in \{1,2,...,n\}$ is given by $p(j) = S(n,j)/B(n)$.

\subsection{Marginal likelihood for one POI}

Suppose we have a set $E$ of $n$ scEPGs, and a single typed POI $s$. If we carry out   the algorithm of \secref{sec:plp} without any trimming of the partitions, then with the  single POI we will end up with a total number of PLPs given by 

\begin{eqnarray*}
	P[n,1] &=& B(n) +\sum_{j=1}^1\sum_{k=1}^n \binom{1}{j}\binom{k}{j} j!\  S(n,k)\\
	&=& B(n) + \sum_{k=1}^n kS(n,k)\\
\end{eqnarray*}	

The first term, $B(n)$, counts the number of the partitions that are not labelled.  The second term counts the labelled partitions: if $n$ scEPGs are partitioned into $k$ non-empty subsets (which happens in $S(n,k)$ ways), then there are $k$ ways in which the $k$ subsets can be labelled with the POI. Or, put another way, for each unlabelled partition having $k$ clusters there are $k$ labelled partitions, corresponding to each cluster being labelled in a partition.

Let $\sigma_{n,k:i}$ denote an unlabelled partition of the $n$ scEPGs having $k$ clusters, its likelihood denoted by $L_{n,k:i}$, with the index $i$ ranging from 1 to $S(n,k)$. Let $p_{n,k:i}$ denote a prior probability that this is the true partition (so that $s$ is not a contributor of any of the cells in $E$). Then we will have

$$p(s \not\in E) = \sum_{k=1}^n \sum_{i=1}^{S(n,k)}p_{n,k:i}$$

Let $\sigma_{n,k:i,j}$ denote a labelled version of the partition $\sigma_{n,k:i}$ , its likelihood denoted by $L^s_{n,k:i,j}$, with the index $j$ ranging from 1 to $k$. Let $p^s_{n,k:i,j}$ denote a prior probability that this is the true partition. Then the probability that $s$ is a contributor of at least one cell in $E$ is 

$$p(s \in E) = \sum_{k=1}^n \sum_{i=1}^{S(n,k)}\sum_{j=1}^k  p^s_{n,k:i,j}$$

with $p(s \not\in E) + p(s \in E) = 1$.

In a court setting it would a priori be assumed that the POI $s$ would have an equal chances to either be or not be a contributor of at least one cell in $E$, that is  $p(s \not\in E) = p(s \in E) = 0.5$. This will happen if we take $p^s_{n,k:i,j} = p_{n,k:i}/k$.

The interpretation of this is as follows. Given that $s$ is not a contributor, then one of the unlabelled partitions is the true one. Giving an equal a priori probability to each of these unlabelled partitions implies a conditional probability for each of $1/B(n)$, agreeing with the case of no POIs.

On the other hand, given that $s$ is a contributor, then the true partition is a labelled partition. Each labelled partition having $k$ clusters is one of $k$ partitions having the same underlying unlabelled partition. Put another way, for each unlabelled partition with $k$ clusters there are $k$ ways of choosing a cluster to label with $s$. If one of these is the true partition, then a priori each have a probability of $1/k$ of being it. Thus
we take  $p^s_{n,k:i,j} = p_{n,k:i}/k$. 

Hence when there is one POI, we apply a factor to each partition equal to the reciprocal of the number of clusters in the partition, arising from the choice of labelling the clusters.

\subsection{Marginal likelihood for two POIs}

The marginal likelihood for two POIs is more complicated, in the sense that there are now several alternative pairs of hypotheses that can be constructed. Let us denote the two POIs by $s$ and $v$, representing the common case that $s$ is a suspect and $v$ a victim in some crime. Then we have the following exclusive possible hypotheses:

\begin{enumerate}
\item[H1:] Neither $s$ nor $v$ contributed at least one  cell to $E$.
\item[H2:] Only  $s$ contributed at least one cell to $E$.	
\item[H3:] Only  $v$ contributed at least one cell to $E$.
\item[H4:] Both  $s$ and $v$ contributed at least one cell to $E$.
\end{enumerate}

The total number of PLPs with 2 POIs is given by

\begin{eqnarray*}
	P[n,2] &=& \sum_{j=0}^2\sum_{k=1}^n \binom{2}{j}\binom{k}{j} j!\  S(n,k)\\
	&=& B(n) + \sum_{k=1}^n \binom{2}{1}\binom{k}{1} 1! S(n,k) 
	+ \sum_{k=1}^n \binom{2}{2}\binom{k}{2} 2! S(n,k) \\
	&=& B(n) + 2 \sum_{k=1}^n kS(n,k) + \sum_{k=2}^n k(k-1)S(n,k)  \\
\end{eqnarray*}

The first term represents the $B(n)$ unlabelled partitions, and corresponds to the terms for $H1$. The second the ways in which the unlabelled partitions  can be labelled with only one of the two labels. The third term represents ways the unlabelled partitions can be labelled with both POIs --- this can only happen if there are at least two clusters.

Let $\sigma_{n,k:i}$ denote an unlabelled partition of the $n$ scEPGs having $k$ clusters, its likelihood denoted by $L_{n,k:i}$, with the index $i$ ranging from 1 to $S(n,k)$. Let $p_{n,k:i}$ denote a prior probability that this is the true partition, and it is not labelled. 

Let  $\sigma^s_{n,k:i,j}$ denote the labelled partition $\sigma_{n,k:i}$  with $s$ labelling the $j$-th cluster, $1 \le j \le k$. Let $p^s_{n,k:i,j}$ denote the prior probability that this is the true labelled partition. Then as the label could have been assigned a priori to any of the $k$ clusters, we shall have

$$p^s_{n,k:i,j} = \frac{p_{n,k:i}}{k}$$

Similarly, if the partition is labelled by $v$ instead of $s$ we will have the corresponding probability

$$p^v_{n,k:i,j} = \frac{p_{n,k:i}}{k}$$

Finally, if the true partition has both POIs, then if $\sigma_{n,k:i}$ is the corresponding de-labelled PLP,  there will be $k(k-1)$ ways of assigning the two distinct labels $s$ and $v$ to the $k$ clusters. We shall have, with the obvious meaning to the notation,

$$p^{s,v}_{n,k:i,j_s, j_v} = \frac{p_{n,k:i}}{k(k-1)}.$$

We thus obtain the following likelihoods for the four labelling hypotheses:

\begin{align*}
	L(H_1) &= C \sum_k \sum_{i=1}^{S(n,k)}L_{n,k:i}\\
	L(H_2) &= C \sum_k \sum_{i=1}^{S(n,k)}\sum_{j=1}^k\frac{L^s_{n,k:i,j}}{k}\\
	L(H_3) &= C \sum_k \sum_{i=1}^{S(n,k)}\sum_{j=1}^k\frac{L^v_{n,k:i,j}}{k}\\
	L(H_4) &= C \sum_k \sum_{i=1}^{S(n,k)}
	\sum_{j^s=1: j^s != j^v}^k\sum_{j^v=1}^k
	\frac{L^{s,v}_{n,k:i,j^s,j^v}}{k(k-1)}\\	
\end{align*}
where in the likelihood for $H_4$ the index $i$ labels the all possible ways of assigning the two labels to the $k$ clusters, and we have set $p_{n,k:i}=C$ a constant for all the unlabelled partitions, thus taking it out as a common factor in each of the expressions of the $L(H_*)$.

The normalizing constant $C$ be found from the constraint that the prior probabilities over the PLPs must sum to 1, thus 

$$\sum_k \sum_i p_{n,k:i}
+ \sum_k \sum_i p^s_{n,k:i}
+  \sum_k \sum_i p^v_{n,k:i}
+  \sum_k \sum_i p^{s,v}_{n,k:i} =1,$$

using  $p^s_{n,k:i}= p^v_{n,k:i} = p_{n,k:i}/k$,
$p^{s,v}_{n,k:i} = p_{n,k:i}/k(k-1)$, and 
setting $p_{n,k:i} = C$. This yields

$$C[B(n) + 2B(n) + B(n)-1] = 1$$

so that

$$C = \frac{1}{4B(n)-1}.$$

However, for calculating odd-ratios the normalization $C$ is unimportant, as the normalization cancels on taking the ratios.

Thus for example

$$\frac{L(H_2)}{L(H_1)} = \frac{\sum_k \sum_{i=1}^{kS(n,k)}{L^s_{n,k:i}}\big/ {k}}
{\sum_k \sum_{i=1}^{S(n,k)}L_{n,k:i}}$$

This is the same as in the previous subsection, obtained using just the single POI $s$.

Similarly, the other odds-ratio of importance is 

$$\frac{L(H_4)}{L(H_3)} = 
\frac{\sum_k \sum_{i=1}^{k(k-1)S(n,k)}{L^{s,v}_{n,k:i}}\big/{k(k-1)}}%
{ \sum_k \sum_{i=1}^{kS(n,k)}{L^v_{n,k:i}}\big/{k}}
$$

Another ratio of interest would be
$$\frac{L(H_2) + L(H_4)}{L(H_1)+L(H_3)} $$
which would be the odds in favour of $s$ being a contributor whether or not $v$ is also a contributor.

\subsection{Marginal likelihood for multiple POIs}

The above patterns extend to more than 2 POIs, and follows from the general formula for the number of PLPs with $m$ POIs:

\begin{eqnarray*}
	P[n,m] &=& \sum_{j=0}^m\sum_{k=1}^n \binom{m}{j}\binom{k}{j} j!\  S(n,k)\\
	&=& B(n) + \sum_{j=1}^m\sum_{k=1}^n \binom{m}{j}\binom{k}{j} j!\  S(n,k) \\
\end{eqnarray*}

For a PLP with exactly $j$ POI labels, we divide the peak height likelihood of the PLP by the factor $\binom{k}{j}j! = k(k-1)(k-2)
..,(k-j+1))$. For a specific set of $j$ POIs selected from the $m$, we add together all the likelihoods for the PLPs that contain only these $j$ POIs, where the likelihoods have been divided by the factor above. On finding odds ratios, the prior normalization constant cancelling out when normalising.

\subsection{Maximum likelihood ratios}

The above has been discussed in terms of marginal likelihoods using  uninformative Bayesian priors over the PLPs. Maximum likelihoods are simply found by taking the highest likelihood term in the marginal likelihood sums. Thus when adjusting the likelihood of a PLP we find the total number of POIs labelling the PLP and the  number of clusters, and divide by the factor $\binom{k}{j}j! $ as above. There is no need to include the prior normalization--- it cancels out when likelihood ratios are evaluated.

\section{Evaluating of partition likelihoods in simulations}

From a computational point of view, it is numerically more stable to work with log-likelihoods rather than likelihoods (the $\log_{10}$ likelihoods of partitions are of the order of negative several thousands). The log-likelihood of a partition is the sum of the log-likelihoods of each of its clusters. When POIs are present, there is a further factor to the likelihood described earlier.

In our likelihood evaluations using the Globafiler\texttrademark\ STR kit, we take the likelihood of a cluster to be the product of the likelihoods evaluated for each individual locus, in which we ignore the data on the Y-linked loci.

If a cluster does not have a POI associated with it, but rather an untyped person, then the likelihood of the cluster on a locus is a sum over all possible genotypes of the $g$ of the form

$$ \sum_g p(g) \prod_c p(c\cd g)$$
where $p(g)$ is the profile probability of the genotype $g$, and the likelihood product of $c$ is over the (peak heights observed in the) scEPGs making up the cluster. 

If a cluster were to be  labelled with a typed person, whether a POI or not, then we would evaluate the likelihood of the cluster conditional on the genotype of the person's profile, that is,  the product $\prod_c p(c\cd g)$ for the  genotype $g$
of the typed person. In our simulations  a typed person is always a POI.

We use a continuous peak height model for evaluating the likelihoods of the peaks observed (or not) in each scEPG. When all scEPGs have been processed, then the likelihoods of the final partitions will be the ones of interest.

 When evaluating peak height likelihoods
the model takes into account the possibilities of drop-in and dropout, instrumental  noise, backward stutters of one or two repeats, and forward stutters of one repeat. For the locus SE33, a locus that has four base pairs in its repeat,  that is prone to high rates of stutter involving 2 base pairs (\ie, half a repeat), the half repeat stutter possibilities are not modelled. Details of the model are omitted here, as we are focussed in this paper on the clustering algorithm, although the quality of the peak height model may be expected to impact on the performance. The model is based on that given in \citep{cowell2018computation,cowell2018unifying}, using a gamma approximation to the tagged  amplicon distribution. Allele frequencies for evaluating genotype probabilities are taken from \cite{butler:etal:03} for autosomal loci of US-Caucasians. An analytic threshold of five RFUs was used throughout.

Suppose that in evaluating the LR of two hypotheses, in the numerator the hypothesis $H_p$ contains a POI $s$, and the denominator the hypothesis $H_d$ has the same contributors as $H_p$ but with $s$ replaced by a random  untyped person from the population. Then it is shown in 
\citep{cowell2015analysis} that the likelihood ratio in favour for a $s$ to be a contributor cannot be greater than the inverse match probability for that person; that is
$$  LR \le 1/MP \to \log(LR) \le -\log(MP) 
\to \frac{\log(LR)}{-\log(MP)} <=1$$
Thus to assess the quality of the likelihood ratios obtained, (under these restricted types of hypotheses pairs),
we divide the log  likelihood ratio by the negative of the log of the match probability of the person. The result is a number with an upper bound of 1; the closer the value to 1 the higher the likelihood that the POI  is a contributor.

Similarly, sub-marginals over PLP likelihoods that have a fixed number $k$ of clusters  could also be evaluated for each possible value of $k$. On normalization, this would yield and estimate of the posterior distribution of the NoC (approximately because of the thinning of PLPs during the search process).

\section{Simulations and their analysis with IPA}

The scEPG admixtures were simulated from the data of nearly 3000 scEPGs used in \cite{grgicak2025dependence}. The scEPGS were split into two groups, a training set of 1947 scEPGs from 43 individuals that were used to calibrate the probability model, and a test set of 996  scEPGs  from 13  individuals used to simulate the admixtures. 
None of the DNA of  individuals in the training set was used  in the testing set.

The scEPGs were amplified using the Globalfiler\texttrademark\  STR assay. In all the analyses in this paper the peaks heights on the Y-linked loci Yindel and DYS391 were not used due to their linkage on the Y-chromosome. 

A number of different simulations were carried out, as detailed in the following
sections. For each we perform various quality checks on the highest likelihood partitions found:

\begin{description}
	\item[NoC errror]
	Let $nc$ denote the number of clusters in the highest likelihood partition; it is the estimate of the number of contributors. We subtract this from the true NoC to give $dc=nc-NoC$; in the ideal case this difference, $dc$, will be zero. 
	\item[Mis-clustering]
	Denoted by $mc$, this counts the number of clusters that have cells from two or more people. Note that we can have $mc \ne 0$ even if $dc = 0$.
	\item[Over-clustering] In this case we count up for each person the number of distinct clusters the person appears in, and we subtract this total (over persons) from  NoC. We denote the result by $oc$. 
\end{description}

If both $dc=0$ and $mc=0$ then we have the correct clustering, and we will then also have $oc = 0$.

The following simple example illustrates the evaluation of these metrics. Let $A_1, A_2$ denote two cells
from person $A$, and $B_1, B_2$ two from person $B$. Then we can have, for example, the following (non exhaustive) clustering possibilities:

\begin{itemize}
	\item $(A_1,A_2)$ and $(B_1,B_2)$. This is the true partition, with $dc = mc = oc = 0$, the ideal case.
	\item  $(A_1)$ and $(A_2,B_1,B_2)$. Here $dc = 0$, as the number of clusters equals the true NoC. There is 1 cluster which has two contributors, so $mc=1$. Person $A$ appears in two clusters, but $B$ in 1, so $oc = 1$. 
	\item  $(A_1,B_1)$ and $(A_2,B_2)$. Here again $dc = 0$. There are two clusters having more than one contributor, so $mc=2$. Both $A$ and $B$ appear in the two clusters, so $oc = 2$. 
	\item  $(A_1)$, $(A_2)$ and $(B_1,B_2)$. Now there are 3 clusters for two contributors, so that $dc = 1$. No cluster has more than one contributor, hence $mc=0$. Contributor $A$ appears in two clusters, but $B$ in 1, so $oc =1 $. 
	\item $(A_1,A_2,B_1,B_2)$, in this case we have just one cluster, hence $dc = -1$. There is one cluster with more than one contributor, hence $mc=1$. Neither $A$ nor $B$ appear in more than one cluster, hence $oc = 0$. 
\end{itemize}

As another illustration, suppose we had a third person $C$ and allow each person to have three cells. Then for following partition $(A_1,B_1,C_1)$,  $(A_2,B_2,C_2)$,  $(A_3,B_3,C_3)$, would have $dc = 0$, as we have three clusters and three contributors. Each of the three clusters has more than one contributors, hence $mc = 3$. Each contributor appears in each of the three cells, giving $oc = 3\times 3 - 3 = 6$.

\subsection{Simulation study 1}

The test set of 996 scEPGs was reduced to the cell scEPGs that had at least 9 allelic peaks that were greater than 20\% of the expected height.
The cells of a test person were retained for sampling if  the individual had at least 25  cells that were retained. This helps to ensure that there is  a good  sample of cells from each person
 in order to  create greater variety in  the simulated admixtures. After this screening, there were 11 test persons and each  had from  between 38 to 99  scEPGs, with a total of  739 cells in all for simulating the admixtures. Using this reduced test dataset, 100 mixtures were simulated  for each of the 11 mixture proportions listed in \tabref{tab:design}, thus making 1100 simulated mixtures in all.

In the next sections we describe results of using the clustering algorithms on these simulated mixtures.  The trimming of the set of partitions after each stage was carried out using the following two algorithms, based on the scores of the partitions and their total number. In each of the 1100 simulations we look at the partition having the highest likelihood.

\begin{enumerate}
	\item If the natural logarithm of  the likelihood of any partition is smaller by 
	$25$ 	
	than the log-likelihood  of the partition having the greatest likelihood, then it is removed. (On a $\log_{10}$ scale the difference is $25/\log(10) \approx 10.8$ .)
	\item If after that there still remain more than 100  partitions, then the 100 highest likelihood partitions are retained, the rest discarded.
\end{enumerate}

\subsubsection{Results for no POIs}

Here we take the simplest case in which we do not use any genotype information of potential contributors nor take any person to be a POI. We are thus using the clustering algorithm of \secref{sec:algnopoi}.
Although we are not using the contributor genotypes in the partition search, we know what they are for each scEPG, and hence can evaluate the quality checks $dc,mc$ and $oc$.

For each of  1100 simulated admixtures, the highest likelihood partition found using IPA was the correct partition except for three of the admixtures.

A closer examination shows that each of these three admixtures  had 1 over-cluster, in each case arising from a true contributions of two cells to the admixtures, and for the same contributor in each case. In all cases there was significant total dropout of alleles and even complete loci, so that there were few loci for which there was overlap of alleles in both cells having non-zero peak heights. 
 However, even on such loci the peaks do not appear to match up in some cases. An example is shown in \tabref{tab:anomaly1}. In the first column we see that the allele 16 peak is quite small. It could be interpreted as either as small peak or as a stutter from the 17 allele with dropout on 16. However, for the peaks in the third column these are more consistent with a genotype of (15,16). The peak on allele 15 here
 can be interpreted as a large stutter from 16, or a drop-in. Either way we have a dropout of 17 which does not occur in the other. With such `clashes' is it perhaps not surprising that the two scEPGs were not put into the same cluster of the highest likelihood partition.
 
\begin{table}[htpb]
	\begin{center}
		\caption{Peak heights on D18S51 for two scEPGs that arose in an overclustering; the genotype of the person generating the two scEPGs is (16,17).\label{tab:anomaly1}}
		\begin{tabular}{c|c|c}
			Allele & Height& Height\\ \hline
			15 & - & 134\\
			16 & 81 & 470 \\
			17 & 392  & - \\
			\hline
		\end{tabular}
	\end{center}
\end{table}

 On one of the three admixtures there appears to be a large clash on two loci, as shown in  \tabref{tab:anomaly2}. Knowing the true genotypes, the peak heights look plausible, with complementary allelic dropouts in both samples. However, without knowing the true genotypes, there is little indication that the contributors have the same genotypes on the two loci. 

\begin{table}[htpb]
	\begin{center}
		\caption{Peak heights on D16S539 and vWA
			 for two scEPGs that arose in an over-clustering. 
			 For D16S539 the true genotype is (9,12), and for
			 vWA 
			  the true genotype is (17,19).\label{tab:anomaly2}}
		\begin{tabular}{c|c|c||c|c|c}
			\multicolumn{3}{c||}{D16S539}&	\multicolumn{3}{c}{vWA}\\ \hline
			Allele & Height& Height & Allele & Height& Height\\ 
			8 & 25 & -  &  16 &- &  41\\
			9 & 480 & - &  17 & - & 228\\
			12 & -  & 231&  18 & 47 & - \\
			&&&				 19 & 423 &- \\
			\hline
		\end{tabular}
	\end{center}
\end{table}

However for this case the peak heights on D1S1656 show greater discordance, as shown in \tabref{tab:anomaly3}: the peaks heights on the second cells suggest a genotype of (14,18) whereas the true genotype is (12,18).

\begin{table}[htpb]
	\begin{center}
		\caption{Peak heights on D1S1656 
			for two scEPGs that arose in an over-clustering. 
			The  true genotype is (12,18).\label{tab:anomaly3}}
		\begin{tabular}{c|c|c}
			Allele & Height& Height \\ \hline
	        11   &   127  &   0 \\
	       12    &  1646  &  0\\
			13   &   0    &   64\\
			14   &   0    &   439\\
			17   &   0    &   48\\
	       18    &  0     &  530\\
			\hline
		\end{tabular}
	\end{center}
\end{table}

These overall results can be compared to those reported in \cite{duffy2023evidentiary} using the MBC algorithm, which state higher error rates. There, 630 mixtures were simulated. With  these,  the MBC algorithm generated partitions with one extra cluster  present in 12\%, 7\%, 7\% and 2\% of respectively the 2,3,4 and 5 persons mixtures. In addition, more than one extra cluster occurred in the partitions generated at rates of 18\%, 8\%, 4\% and 3\% respectively of the 2,3,4 and 5 person mixtures. (A direct comparison with the MBC would require running the algorithm on the same set of simulated mixtures, and using their same peak height model, so these values should be treated with care in drawing firm conclusions about the comparing the performance of each clustering algorithm.)

\subsubsection{Results with POIs}

The setup for these is similar to that of the non-POI simulation. 
We use the same 739 cells from 11 persons as for the non-POI simulation. But now we create a for each of these 11 persons 100 mixtures for each design (1100 admixtures for each person) such that a POI is forced to be a contributor to the admixture.
Thus we have $11 \times 1100 = 12,100 $ admixtures in total.

Then for each person we do the POI-clustering algorithm of 
\secref{sec:algpoi} treating the person as a POI with known genotype. On clustering a particular admixture we have a set of PLPs. We search these to find the highest likelihood PLP that has the person as a specified contributor to one of the subsets of the partition. The cluster is then examined for mis-clustering and over-clustering.

\textit{In summary, there was no mis-clustering in any of the 12,100 admixtures. There were 11 admixtures for which there was a single over-clustering. }

\subsection{Simulation study 2}

Because of the almost perfect results of the clustering algorithm of this paper reported above, a further simulation was carried out.
In this 400 admixtures were generated in which 
4 cells were sampled for each of 7  contributors randomly selected from the 11 test persons used in the two earlier simulations. 
On performing a partition search on each of these without a POI, it was found that in every case the highest likelihood partition 
was a correct clustering with neither mis-clusterings nor over-clusterings.

\section{Hierarchical agglomerative clustering}
\label{sec:aggloclust}

\cite{lun2024calculation} proposed an algorithm based on \textit{hierarchical  clustering}, which 
uses the same peak height likelihood model to evaluate partition likelihoods as used in the clustering, like the  IPA of this paper, and unlike the MBC approach which uses a pattern matching algorithm for the clustering. 
They called their algorithm \textit{forensic-aware clustering (FAC)}. A comparison of MBC and FAC was recently carried out by 
\cite{qhawe2025} using admixtures simulated from scEPGs. Here we shall compare FAC with IPA using a similar set of simulated admixtures.

\subsection{A choice of priors for FAC}
 
The method of clustering used in the FAC method is \textit{hierarchical agglomerative clustering} (HAC) which works as follows.

Suppose we have a set of $n$ scEPGs. We make an initial partition in which each scEPG is in its own cluster, and we calculate the likelihood of this partition. We then repeatedly find the pair of distinct clusters in the partition which, if combined, would make  the greatest increase in the partition likelihood, and so combine them to form a new partition with one fewer cluster. We repeat this until there is no pair of clusters which when combined would increase the partition likelihood, and then the  clustering algorithm  halts. However there is here a subtle choice to be made in prior assumptions, as we now discuss.

Suppose we have $m$ scEPGs to  cluster. Then the NoC could be any integer
in the range $[1,m]$.   \cite{lun2024calculation} propose for their method that each of these $m$ integer values is equally likely, and further given the number of clusters is $k$, that each partition with $k$ clusters is assigned  equal \textit{a prior} probability. This implies that the prior distribution associated with a partition of $k$ clusters is proportional to the inverse of the Stirling number of the second kind $S(m,k)$

Now consider the FAC has reduced  the number of clusters to $k$, and let ${C_i: i \in \{1,\cdots,k\}}$ denote the $k$ clusters. Let $L(C_i)$ and $L(C_j)$ denote the likelihoods of two distinct clusters, and $L(C_i,C_j)$ denote the likelihood if the clusters were combined into 1, in which the partition prior has not been incorporated. Then the FAC evaluates for each pair of clusters  $1 \le  i < j \le k$ the following quantity:

$$\delta_{i,j} = \log\frac{L(C_i,C_j)}{S(m,k-1)} -
	\log\frac{L(C_i)L(C_j)}{S(m,k)}.$$
	
which uses Stirling numbers to scale the likelihoods, as dictated bvy the choice of a uniform prior on the NoC. We use $S(m,k-1)$ in the first term because if we combine the two clusters, then we are left with $k-1$ clusters in the partition, and there are $S(m,k-1)$ ways of partitioning the $m$ cells into $k-1$ non-empty clusters. The second term uses $S(m,k)$ because we still have $k$ clusters if we do not combine the two. 

The FAC algorithm combines the pair of clusters  that maximizes $\delta_{i,j}$, provided that this quantity is positive. If it is $\le 0$ then the FAC algorithm stops attempting to combine clusters.

The choice of prior is different in the IPA, in which we put a uniform prior over all the partitions.  So, when adding a cell we may simply compute the  log-likelihood of a partition as the sum of the log-likelihoods of its constituent clusters. 

If we were to modify FAC to use a uniform prior on partitions, rather than NoC, then the Stirling numbers would not appear in the formula for $\delta_{i,j}$, and we would have instead the following modification which is in line with that used in IPA:

$$\delta^*_{i,j} = \log L(C_i,C_j) -\log L(C_i) - \log L(C_j).$$

The  algorithm would the combine clusters that maximizes $\delta^*_{i,j}$ provided that the maximum value is positive; otherwise the algorithm terminates.

The difference in these two quantities is simply

$$\delta_{i,j}  - \delta^*_{i,j} = \log\frac{S(m,k)}{S(m,k-1)}.$$

In comparing IPA with FAC, we shall consider the two variant algorithms using both possible priors described above. That using $\delta_{i,j}$ will  be called FAC1, and that  using $\delta^*_{i,j}$ will  be called FAC2. The term HAC will refer to either FAC1 and FAC2.

In using a uniform prior over partitions, we implicitly imply a non-uniform prior over the NoC. Given $n$ scEPGs we will have the prior for $k\in [1,n]$ contributors to be the ratio of Stirling numbers of the second kind to Bell number $B(n)$: 

$$ P(k) = \frac{S(n,k)}{B(n)}$$

The following plot shows this prior distribution for the case that the number of scEPGs is given by $n=20$. We see from the plot that the distribution favours lower $k$ values; the mean of the distribution is approximately 8.18, which is smaller than that of a uniform prior (over NoC) mean value of 11.5.

\includegraphics[scale=0.3]{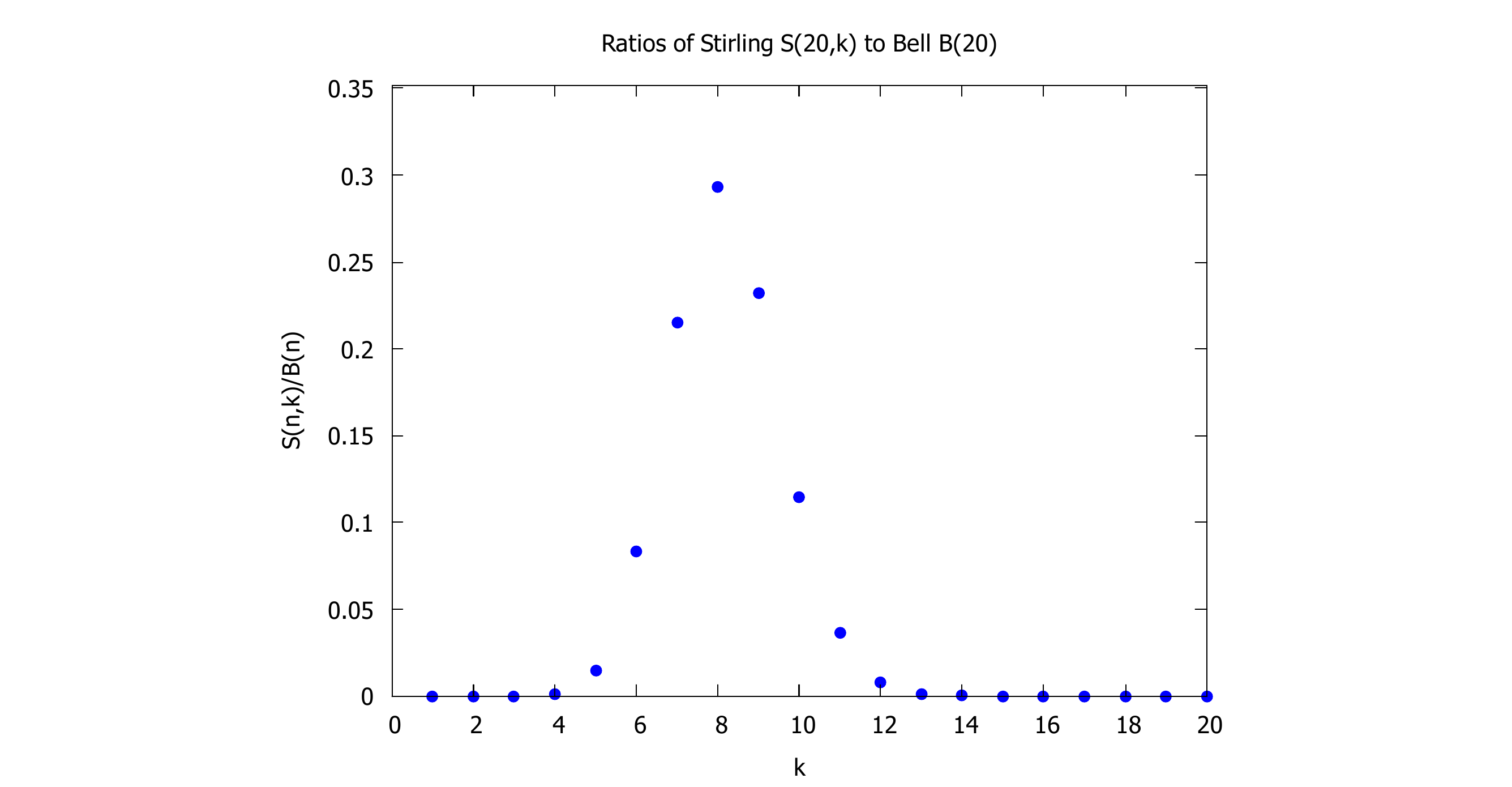}

\subsection{Comparison of HAC and IPA clustering: theory}

One major difference of HAC to IPA  is that, at the end of the HAC algorithm there is only one partition. With the results of the IPA,  having not just the highest likelihood partition found by the search means that a posterior distribution may be found for the retained partitions (using for example a prior assumption that all partitions are equally likely). If the highest likelihood partition has a very much higher posterior probability than the second most likely partition and those later in the sequence, then the inferences based on it alone will garner high confidence. On the other hand if this is not the case, then a Bayesian averaging over the retained partitions would be more appropriate.  The HAC algorithm, giving just one final partition, does not permit this, and there is no easy way to judge the confidence of the partition found in any particular case. 

Another aspect is that if at an intermediate stage of the HAC algorithm a mis-clustering  occurs, then it cannot be undone by later further clusterings; any such cluster will occur as a subset of the scEPGs of one of the clusters in the final partition. This is not necessarily the case for the IPA, although this will depend on the degree to which the partitions are trimmed after each cell is added.

A large difference is in the treatment of typed POIs, and in the evaluation of likelihood ratios for a POI to be in a mixture versus they not being in the mixture.  In the case of HAC, the partition is found without using the type information of the POI. If we index the clusters by $i$, then each cluster $C_i$ has associated with an assumed contributor $u_i$ of unknown genotype $g_i$ on locus $m$. The partition likelihood is then (schematically)

$$ \prod_i \left( \sum_{g_i} P(C_i\vert u_i, g_i)p(g_i) \right)$$

To calculate the likelihood for the POI, the partition is assumed to be correct, and one amends the above sum by averaging the likelihoods obtained by restricting each $u_i$ in turn to be the POI, so that the sum over the $g_i$ is simply replaced by the term matching  the genotype of the  POI, and also omitting the $p(g)$ factor. The ratios of these likelihood may then be obtained for the LR.

In contrast IPA has the modification presented above in terms of partially labelled partitions. The likelihood for the POI to be in the set of scEPG cells is then taken to be either (i) the likelihood of the first partition to have a cluster labelled by the POI, or (ii), in a Bayesian manner, take the sum of the posterior probabilities of the partitions that have a cluster labelled by the POI. For the denominator of the LR take for (i) the highest likelihood partition that does not have the POI labelling a cluster; or for (ii) find the sum of posterior distributions of such partitions. Taking the ratio for (i) gives the likelihood ratio, whilst for (ii) gives the Bayesian posterior odds.

There is another subtlety here for the maximum likelihoods evaluated for (i).  It is not the case that the highest likelihood PLP with the POI will have the same set of clusters as that of the PLP without the POI - the two partitions could even have  different numbers of clusters. When this happens the LR ratio is not then of the standard form given by:  the likelihood of the POI and a set number $k-1$ of untyped persons being the contributors versus another that of  $k$ untyped persons being the contributors. 

From a computational point of view the complexity of the algorithms is quite different. Suppose that there are $n$ scEPGs to cluster. For  HAC  the first step is to evaluate all $n$ individual cell likelihoods,  and the likelihoods  for the $n(n-1)/2$  distinct possible cluster pairs. The highest scoring is chosen. Then the $n-2$ likelihoods of combining this cluster with each of the remaining singleton clusters is found. This process is continued, and we see that the complexity is $O(n^2)$. 

Less can be said about IPA, as the complexity will depend upon the trimming criterion.  In the worst case trimming is not applied, and so the complexity would be $O(B(n))$ (in the case of no POIs), the number of possible unlabelled partitions. At the other extreme if only  the highest likelihood is  retained  after a cell is added then the complexity will be $O(n)$. In between these extremes, if one retains at most $M$ partitions after trimming then the complexity will be 
$O(nM)$. But the trimming also depends upon the difference between a partition and the highest likelihood partition at each stage. It turns out that when the scEPGs are of good quality then this difference is large and few partitions ar retained, so the algorithm in that case is also approximately $O(n)$.

\subsection{Comparison of HAC and IPA: simulations}

We now turn attention to how the two algorithms compare in practice. We discuss several simulations.

\subsubsection{Simulation study 4}

In this  we reuse the 1100 admixtures of simulation 1, based upon the designs in \tabref{tab:design}. 

As discussed earlier, for all but three admixtures the most likely partition return by IPA was the correct partition.
IPA and FAC2 found the same partition for all 1100 admixtures. 
In the case of FAC1, there were 17 admixtures for which a different partition was found from IPA (and FAC1); none of these were the three admixtures for which IPA did not identify the correct partition.

In 15 of the 17 admixtures there was a single over-clustering,
with a cluster of 2 split into two singleton clusters. In the other two cases, there was a single mis-clustering, in which one of the cells of a two-cell cluster was grouped with another cluster.

\subsubsection{Simulation study 5}

In this we reuse the 400 simulations of admixtures of 7 persons with 4 cells each of simulation 2. For all admixtures IPA, FAC1 and FAC2  returned the same highest likelihood partitions, which were also the ground truth true partitions.

\subsection{Simulation study 6}

The previous two comparisons were based on samples of scEPGs that had at least nine relatively large peaks. As these are of reasonable quality, that could explain the good results and agreements between the  clustering algorithms. In this simulation we draw 500
 admixtures  in  which there are 5 persons each contributing 15 cells, that is the design is $\{15,15,15,15,15\}$ having  75 cells per admixture. We restrict the 
 cells to be sampled  to those having at least 2 and no more than 30 large peaks. This therefore admits many more low quality scEPGs.
 This lead to the cells being sampled from 6 persons each having from between 26 to 53 cells to sample from.

For these mixtures the FAC2  and IPA again agreed on the same clustering for the most likely partition on each admixture. However, 
only 61 admixtures were correctly partitioned. There were 330 admixtures for which FAC1 disagreed with IPA for the most likely partition. Of these FAC1 found the true partition in 32 cases, whilst IPA had either 1 or 2 over-clusterings, (an none of the 32 admixtures had mis-clusterings)

The results for IPA and FAC2  are summarised in the following table, with mc representing mis-clusters, and oc over-clusters. We see there is a tendency to greater over clustering compared to mis-clustering.

\begin{center}

\begin{tabular}{c|cccccc|c}
	& \multicolumn{6}{|c}{oc} \\
	mc & 0&1& 2 & 3 &4&5& total\\ \hline
	0 & 61	&104	&34	&3	&0&	0 &202\\
	1 &0 &72	&101&	52&	9	&0&234\\
	2 & 0&	0	&26	&25	&11&	2&64\\ \hline
	total&61&176&161&80&20&2&500\\
\end{tabular}
\end{center}

The results for FAC1 are in the following table, which although showing approximately 50\% more correct partitions also exhibits greater mis-classifications for those incorrectly identified.

\begin{center}
	\begin{tabular}{c|cccccc|c}
			& \multicolumn{6}{|c|}{oc} \\
		mc & 0 & 1 & 2 & 3 & 4 & 5 & total  \\ \hline
		0 & 92 & 6 & 0 & 0 & 0 & 0 & 98  \\
		1 & 0 & 187 & 11 & 1 & 1 & 0 & 200  \\
		2 & 0 & 0 & 147 & 11 & 1 & 1 & 160  \\
		3 & 0 & 0 & 0 & 38 & 1 & 0 & 39  \\\hline
		total & 92 & 193 & 158 & 50 & 6 & 1 & 500  \\
		\hline
	\end{tabular}
\end{center}

\subsection{Simulation study 7}

In these simulations we take the same cells as used in the  simulation 6 (which had some cells of low quality), but now sample, for each of the six persons,  100 admixtures each of the  11 designs in \tabref{tab:design}
such that the person is forced to be a contributor (making 6600 admixtures in all). As expected there were a large number of admixtures which were incorrectly
clustered as judged by the highest likelihood partitions, from between 496 to 674 depending on the person forced to be in the admixture.

The FAC1 algorithm gave differences to IPA from between 590 and 728 admixtures in the depending on the person forced to be in the admixture, with a total of 3583 over all admixtures, a little over 58\%.

In contrast the FAC2 algorithm differed from IPA in its selection of the highest likelihood partition for 10 admixtures  10 out of the 6600 admixtures.
In all these 10 cases the maximum likelihood found by IPA was greater than that found by FAC2, but only by a small amount, ranging from 0.0076 to 0.2913 on the $\log_{10}$ scale, that is, factors from 1.02 to 1.96.

\section{Distributions of Log-likelihood ratios}

\subsection{Good quality scEPGs: simulation 2 revisited}

We begin with  the (higher quality) scEPG admixtures sampled in Simulation 2 based on the designs of \tabref{tab:design}. Recall that
we created  for each of these 11 persons 100 mixtures for each design (1100 admixtures for each person) such that each person is forced to be a contributor to the 100 admixtures of each design, admixture, leading to 1,100  admixtures for each person for which they are known to be a contributor, and 12,100 such mixtures in all for the 11 persons.  

\begin{figure}[htbp]
	\includegraphics[scale=0.8]{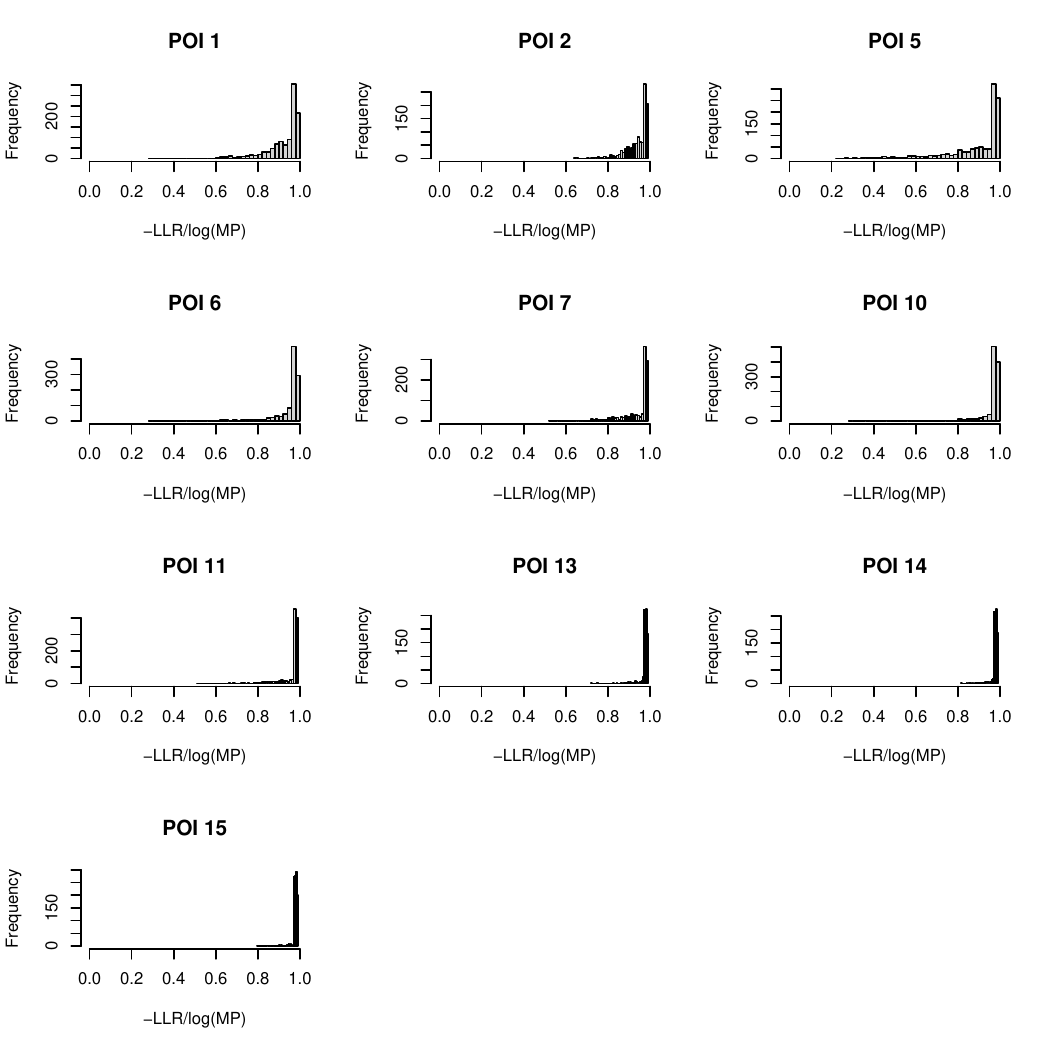}
	\caption{Log-likelihood ratios in favour of a POI being a contributor,  normalized by the negative log match probability for each set simulations of high quality admixtures for  each of the 10 POIs.\label{fig:llrsgood}}
\end{figure}

On a log10 scale, the match probabilities of the  POIs ranged from -29.2677 to -26.7071.  \figref{fig:llrsgood}
shows the  log-likelihood ratios normalized using the log match probabilities for each of the ten POIs.
We see that most of the values are greater than 0.8, with many close to 1. A value of 0.3 would correspond to a likelihood ratio in favour of a person matching the genotype of the POI of
$10^{0.3\times26.7}\approx 10^{8}$.  There are a total of six ratios less than 0.3, the smallest being 0.231 corresponding to a likelihood ratio of approximately $10^6$.

\clearpage

\subsection{Poor quality scEPGs: simulation 8}

For these we repeat the generation of admixtures as carried out in simulation 2, but now using the poorer quality cells used in  simulation 6 for which there  were 6 POIs. The match probabilities of the six   POIs ranged from -29.2677 to -27.4782. Frequency histograms 
of the  $-LLR/\log(MP)$ ratios for each POI are shown in the following figure. \figref{fig:llrspoor}
shows the normalized log-likelihood ratios for each of the six POIs.
We see the distributions are more spread out compared to the good  quality admixtures. Nevertheless, most values are greater than 0.3, corresponding to a likelihood ratio in favour  of $10^8$. There are 343 values less that 0.3. Again, an examination of these admixtures showed that the POI contributed just two cells in each case. There are six negative values, ranging from -0.065 to -0.002. This is similar to the findings in \cite{grgicak2025dependence} 
of negative LLRs obtained with using just a single cell when 
evaluating likelihoods of low quality cells.

	\begin{figure}[htbp]
		\includegraphics[scale=0.8]{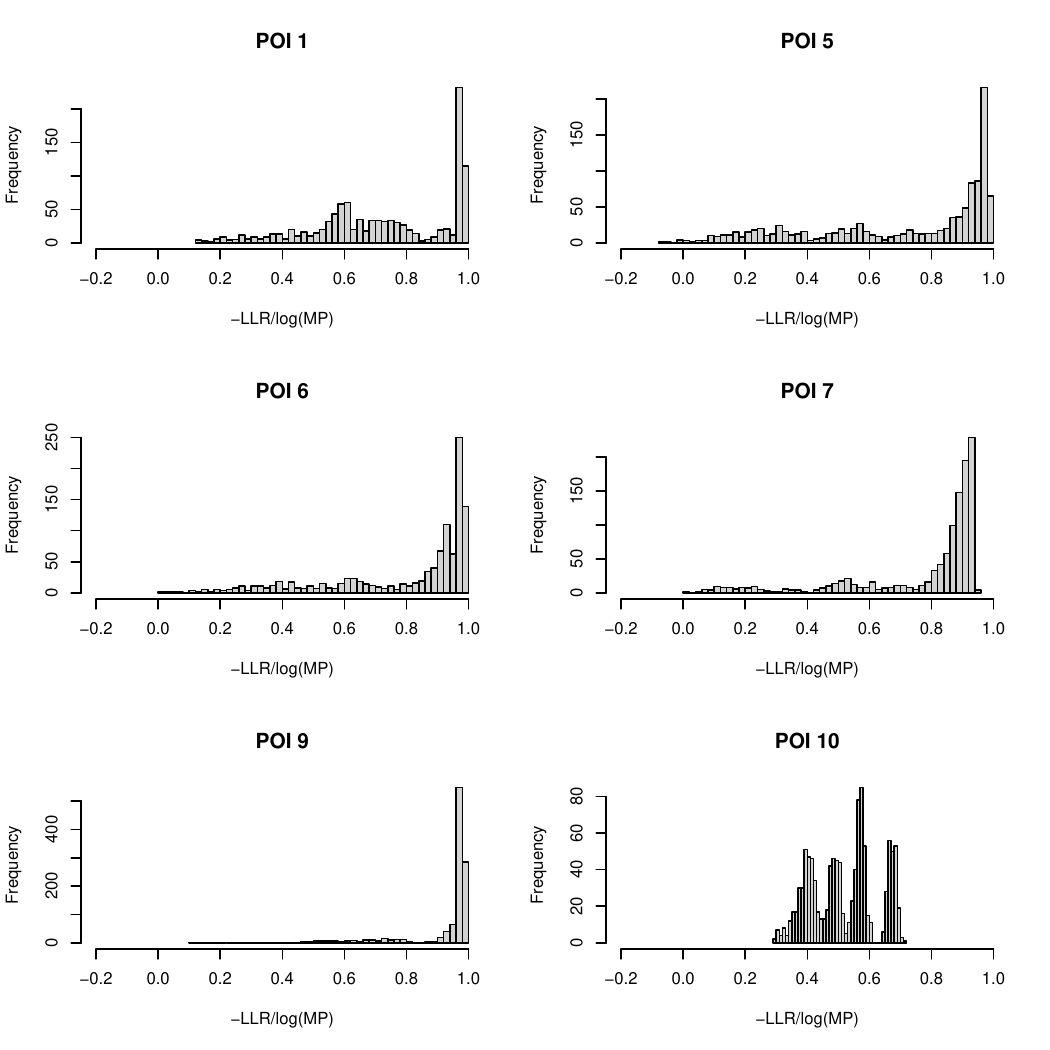}
		\caption{Log-likelihood ratios in favour of a POI being a contributor,  normalized by the negative log match probability for each set simulations of lower quality admixtures for  each of the 6 POIs.\label{fig:llrspoor}}
	\end{figure}
	
\clearpage

\section{Summary and extensions}

\subsection{Summary}
A simple probabilistic clustering algorithm for grouping scEPGs, that could be generated from DNA mixtures, into clusters of cells from  individual contributors has been presented. The algorithm appears to be highly effective, with very low error rates if the scEPGS are of good quality showing many high peaks, for both variations of no POIs begin specified and with POIs.

A comparison was made with hierarchical agglomerative clustering algorithms. Two variations of HAC were considered, one with a uniform prior on the number of contributors, as proposed by  \cite{lun2024calculation} 
(which we denote by FAC1), and one which uses a uniform prior over partitions (which we denote by FAC2).  The IPA of this paper almost always agrees with the highest likelihood partition found by the FAC2 algorithm, with lower error rates for good quality cells. An advantage of the IPA algorithm is that it generates a set of high likelihood partitions, whereas HAC algorithms generate just one. If there is a large difference in the likelihoods between the highest and second highest likelihood partitions found by IPA, then this lends confidence that the highest likelihood partition is the true partition.

We conclude that the search procedure given in  this paper could be a valuable tool in making the future use of single-cell isolation and  amplification a viable and productive method for analysing DNA mixtures. This is particularly the case for mixtures of many persons, where current probabilistic genotyping software has a computational complexity exponential in the number of contributors, whereas the clustering algorithm of this paper has a complexity approximate linear in the number of scEPGs. The results from simulations 2 and 5  are particularly pertinent here, involving seven contributors donating equal numbers of cells. Current bulk mixture probabilistic genotyping software is limited to around 5 possible contributors in order for computations to be manageable. In addition, bulk mixture analysis is dependent upon people contributing quite different amounts of DNA from each other to identify them, whereas in these simulations all contributors donated the same amount of DNA, something that bulk mixture programs have a very hard time dealing with.

The performance of the algorithm depends upon the quality of the scEPG. When we retained only scEPGs having 9 or more peaks which were 20\% of greater than the mean allelic peak height, almost no mis-clusterings resulted from IPA and FAC2. Allowing lower quality scEPGs into the admixtures led, not unexpectedly, to mis-clustering and over-clustering becoming more frequent. 

\subsection{Extensions}

In bulk mixture analysis  replications --- in which several subsamples of a  DNA mixture amplified separately --- are sometimes used to analyse a DNA mixture when a single amplification is not greatly informative, 
for example in low template samples. For the single cell analysis of this paper, the situation is different. Should a set of scEPGs from a sample not be informative enough, then more scEPGs could be made from another sampling of the DNA mixture. In this case it would make sense to combine the two sets of scEPGs for a single larger partition search. This would be particularly appropriate in cases where the cluster of a partition associated with a POI has just a few cells: combining the replicates may be expected to increase the number of cells from the POI \textit{if} the POI is a contributor to the DNA mixture.

In bulk mixture analysis,  replicates can be amplified using different kits. The IPA  of this paper can be extended to this case quite straightforwardly- when adding a new cell amplified by a kit, use only the loci of the kit for modifying the likelihood. This will be most useful when the kits have a large overlap of loci, and especially useful when considering POIs that have been profiled with the kits used. 

The extension of IPA to handle multiple samples is also relatively straightforward.  If we have two distinct DNA samples recovered at a crime scene, they might have only a partial overlap of common contributors, and perhaps none at all. A POI might be in one but not the other. However, combining the two sets of scEPGS that might be obtained from the two samples is still possible - one then tries to find the total number of distinct persons contributing to the two DNA mixtures. One would expect that cells originating from a common contributor to both samples would have their cells clustered by the IPA applied to the combined set of scEPGs. The likelihood ratio for a POI would be in terms of the likelihood for the POI to be in at least one sample. In addition the IPA can  also be carried out on each set of scEPGs separately, giving individual likelihood ratios for a POI to be in each sample separately.

One assumption used in this paper is that there is no known relatedness
between the persons making up the assumed mixture of scEPGs. At present no suggestions for modifying the IPA are given here; it would be interesting to modify IPA to include  relatedness assumptions. Similarly, it would be useful to overcome the approximation made in applying substructure corrections to the likelihoods.

\section*{Acknowledgements} 
The  author would like to thank Catherine Grgicak and Desmond Lun for  discussions about, and  confidential access to, 
the single cell data used in these simulations.  Plots were produced using gnuplot \citep{gnuplot}, R \citep{R} and wxMaxima \citep{wxmaxima}.

\cleardoublepage 
\addcontentsline{toc}{section}{\bibname}
\bibliographystyle{abbrvnat} \markboth{\bibname}{\bibname}
\bibliography{singlecellsim}

\end{document}